\documentclass[superscriptaddress,aps,preprintnumbers,amsmath,showpacs,amssymb,prd,nofootinbib,reprint]{revtex4-1}

\usepackage[dvipdfmx]{graphicx}
\usepackage{amsfonts}
\usepackage[mathscr]{eucal}
\usepackage{ascmac}
\usepackage{dcolumn}
\usepackage{bm}
\usepackage[colorlinks=true,linkcolor=blue,citecolor=blue]{hyperref}
\usepackage{hyperref}
\usepackage{color}

\begin{document}


\newcommand{\vev}[1]{ \left\langle {#1} \right\rangle }
\newcommand{\bra}[1]{ \langle {#1} | }
\newcommand{\ket}[1]{ | {#1} \rangle }
\newcommand{\eV}{ \ {\rm eV} }
\newcommand{\KeV}{ \ {\rm keV} }
\newcommand{\MeV}{\  {\rm MeV} }
\newcommand{\GeV}{\  {\rm GeV} }
\newcommand{\TeV}{\  {\rm TeV} }
\newcommand{\1}{\mbox{1}\hspace{-0.25em}\mbox{l}}
\newcommand{\Red}[1]{{\color{red} {#1}}}

\newcommand{\lmk}{\left(}  
\newcommand{\rmk}{\right)}
\newcommand{\lkk}{\left[}  
\newcommand{\rkk}{\right]}
\newcommand{\lhk}{\left \{ }  
\newcommand{\rhk}{\right \} }
\newcommand{\del}{\partial}  
\newcommand{\la}{\left\langle} 
\newcommand{\ra}{\right\rangle}
\newcommand{\half}{\frac{1}{2}}

\newcommand{\bea}{\begin{array}}
\newcommand{\eea}{\end{array}}
\newcommand{\beq}{\begin{eqnarray}}
\newcommand{\eeq}{\end{eqnarray}}

\newcommand{\dd}{\mathrm{d}}
\newcommand{\Mpl}{M_{\rm Pl}}
\newcommand{\mg}{m_{3/2}}
\newcommand{\abs}[1]{\left\vert {#1} \right\vert}
\newcommand{\mphi}{m_{\phi}}
\newcommand{\Hz}{\ {\rm Hz}}
\newcommand{\for}{\quad \text{for }}
\newcommand{\Min}{\text{Min}}
\newcommand{\Max}{\text{Max}}
\newcommand{\Kahler}{K\"{a}hler }
\newcommand{\cphi}{\varphi}
\newcommand{\Tr}{\text{Tr}}
\newcommand{\diag}{{\rm diag}}

\newcommand{\SUf}{SU(3)_{\rm f}}
\newcommand{\Upq}{U(1)_{\rm PQ}}
\newcommand{\Zpq}{Z^{\rm PQ}_3}
\newcommand{\Cpq}{C_{\rm PQ}}
\newcommand{\ubar}{u^c}
\newcommand{\dbar}{d^c}
\newcommand{\ebar}{e^c}
\newcommand{\nubar}{\nu^c}
\newcommand{\Ndw}{N_{\rm DW}}
\newcommand{\Fpq}{F_{\rm PQ}}
\newcommand{\fpq}{v_{\rm PQ}}
\newcommand{\Br}{{\rm Br}}
\newcommand{\Lag}{\mathcal{L}}
\newcommand{\Lqcd}{\Lambda_{\rm QCD}}
\newcommand{\const}{\text{const}}

\newcommand{\ji}{j_{\rm inf}} 
\newcommand{\jb}{j_{B-L}} 
\newcommand{\M}{M} 
\newcommand{\im}{{\rm Im} }
\newcommand{\re}{{\rm Re} }
\newcommand{\cm}{\ {\rm cm} }

\def\lrf#1#2{ \left(\frac{#1}{#2}\right)}
\def\lrfp#1#2#3{ \left(\frac{#1}{#2} \right)^{#3}}
\def\lrp#1#2{\left( #1 \right)^{#2}}
\def\REF#1{Ref.~\cite{#1}}
\def\SEC#1{Sec.~\ref{#1}}
\def\FIG#1{Fig.~\ref{#1}}
\def\EQ#1{Eq.~(\ref{#1})}
\def\EQS#1{Eqs.~(\ref{#1})}
\def\blue#1{\textcolor{blue}{#1}}
\def\red#1{\textcolor{blue}{#1}}

\newcommand{\fa}{f_{a}}
\newcommand{\Uh}{U(1)$_{\rm H}$}
\newcommand{\osc}{_{\rm osc}}

\newcommand{\mav}{\left. m_a^2 \right\vert_{T=0}}
\newcommand{\mat}{m_{a, {\rm QCD}}^2 (T)}
\newcommand{\mam}{m_{a, {\rm M}}^2 }
\def\eq#1{Eq.~(\ref{#1})}

\newcommand{\LQCD}{\Lambda_{\rm QCD}}

\newcommand{\UH}{U(1)$_H$ }

\newcommand{\EV}{ \ {\rm eV} }
\newcommand{\KEV}{ \ {\rm keV} }
\newcommand{\MEV}{\  {\rm MeV} }
\newcommand{\GEV}{\  {\rm GeV} }
\newcommand{\TEV}{\  {\rm TeV} }

\def\order#1{\mathcal{O}(#1)}


\preprint{
IPMU 16-0080
}

\title{
SIMP from a strong U(1) gauge theory 
with a monopole condensation 
}

\author{
Ayuki Kamada
}
\affiliation{
Department of Physics and Astronomy, 
University of California, Riverside, CA 92521, USA 
}

\author{
Masaki Yamada
}
\affiliation{Department of Physics, Tohoku University, 
Sendai, Miyagi 980-8578, Japan} 

\author{
Tsutomu T. Yanagida
}
\affiliation{Kavli IPMU (WPI), UTIAS, 
The University of Tokyo, 
Kashiwa, Chiba 277-8583, Japan}

\author{
Kazuya Yonekura
}
\affiliation{Kavli IPMU (WPI), UTIAS, 
The University of Tokyo, 
Kashiwa, Chiba 277-8583, Japan}

\date{\today}

\begin{abstract} 
We provide a variant model of strongly interacting massive particle (SIMP), 
where 
composite dark matter comes from 
a strongly interacting U(1) theory. 
We first explain a non-Abelian version of the model with 
an additional singlet field, 
which is mixed with the Higgs field to maintain the kinetic equilibrium between 
the hidden and Standard Model (SM) sectors. 
The mixing leads to signals 
that 
would be detected by future collider experiments, 
direct DM detection experiments, and beam-dump experiments. 
Then we investigate a U(1) theory with a scalar monopole, 
where U(1) charged particles are confined by monopole condensation. 
In this model, 
the radial component of monopole can mix with the Higgs field, 
so that we do not need to introduce the additional singlet field.

\end{abstract}

\maketitle

\section{Introduction
\label{sec:introduction}}

Strongly interacting confining theories 
are widely considered as a physics beyond the Standard Model (SM). 
In the literature, 
they focused on non-Abelian gauge theories 
to realize strong interactions and confinement because 
they are asymptotic free and become strong at low energy. 
However, 
Abelian gauge theories can realize confinement 
by a monopole condensation, 
which was 
discussed to aim to undestand the confinement of QCD~\cite{Nambu:1974zg}. 
There are many theoretical studies to reveal properties of 
theories with electrons and monopoles in $\mathcal{N}=2$~\cite{Argyres:1995jj, Argyres:1995xn} 
and $\mathcal{N}=1$ supersymmetry~\cite{Bolognesi:2015wta, Giacomelli:2014rna, Xie:2016hny, Buican:2016hnq}. 
In cosmology, confining U(1) theories 
have some advantage compared with non-Abelian models. 
For example, 
a particle content is economical~\cite{Yamada:2015waa} 
and 
there is no baryon state at low energy~\cite{Yamada:2016jgg}.

In the literature, 
a strongly interacting massive particle (SIMP) 
is widely considered 
as a candidate for dark matter (DM) 
because its self-interaction 
can address some astrophysical offset 
such as "core v.s. cusp" problem 
and "too-big-to-fail" problem~\cite{Spergel:1999mh, deBlok:2009sp, BoylanKolchin:2011de, Zavala:2012us}. 
A recent observation of Abell3827 cluster 
also favours SIMP DM models~\cite{Kahlhoefer:2015vua} (see also Ref.~\cite{Massey:2015dkw}). 
It has been pointed out that 
$\order{100} \MeV$ DM 
can have a observed relic density 
by the freezeout mechanism via $3 \to 2$ annihilation process 
and have a correct scattering cross section required to explain the above 
astrophysical problems and observations~\cite{Hochberg:2014dra}. 
In particular, 
such a model can be realized by 
a low energy effective theory 
of a strongly interacting non-Abelian gauge theory~\cite{Hochberg:2014kqa}. 
The SIMP models 
require that 
the DM sector is in kinetic equilibrium with the SM sector 
so that 
the DM does not become hot via $3 \to 2$ annihilation process. 
In order to realize the kinetic equilibrium, 
they may introduce a hidden U(1) gauge boson 
which has a small kinetic mixing with SM U(1)$_Y$ gauge boson~\cite{Lee:2015gsa, Hochberg:2015vrg}.

In this paper, 
we provide simple models to realize the SIMP mechanism 
that predicts correct relic DM abundance 
and scattering cross section indicated by astrophysical observations. 
First we consider a non-Abelian gauge theory 
with a singlet field in the hidden sector. 
Assuming a mixing between the singlet field 
and the SM Higgs field, 
we can maintain the kinetic equilibrium between the hidden and SM sectors. 
The mixing effect leads to 
signals 
for future collider experiments, 
direct DM detection experiments, and beam-dump experiments. 
Then 
we provide a SIMP model in an Abelian gauge theory 
that is confined due to a monopole condensation. 
In this model, 
the radial component of monopole plays the role of the singlet field of the former model, 
which naturally realize the kinetic equilibrium between the hidden and SM sectors. 
It is outstanding that 
the monopole plays the roles of U(1) confinement and mediator between the hidden and SM sectors.

\section{SIMP with a singlet field
\label{model}}

\subsection{SIMP and its thermal relic 
\label{sec2-2}}

First, we consider a variant SIMP model in a non-Abelian gauge theory. 
We introduce a singlet field $S$ with $N_F$ pairs of hidden quarks $Q_i$ and $\bar{Q}_i$, 
which are charged under a hidden SU($N$) gauge symmetry 
in the fundamental and anti-fundamental representation, respectively. 
The hidden quarks interact with the singlet field via the following Yukawa interaction: 
\beq
 \mathcal{L}_{\rm int} = \lambda S Q_i \bar{Q}_i + {\rm h.c.}, 
 \label{singlet}
\eeq
where we assume SU($N_F$)$_V$ flavour symmetry (see Table~\ref{table1}). 
As we discuss in Sec.~\ref{sec2-3}, 
we consider the case that 
the singlet field has a nonzero vacuum expectation value (VEV), 
which gives a mass for hidden quarks such as $m_Q = \lambda \la S \ra$. 
We also assume that 
there is a mixing between the singlet field $S$ and the SM Higgs field $h$ 
with a mixing parameter $\theta$, 
which gives interactions between the hidden and SM sectors.

\begin{table}\begin{center}
\begin{tabular}{|p{1.0cm}|p{1.5cm}|p{1.5cm}|p{1.5cm}|p{1.5cm}|}
  \hline
  \rule[-5pt]{0pt}{15pt}
& \hfil SU($N_F$)$_V$ \hfil 
    & \hfil SU($N$) \hfil \\
  \hline
  \rule[-5pt]{0pt}{15pt}
  \hfil $Q_i$ \hfil 
& \hfil $\Box$ \hfil 
  & \hfil $\Box$ \hfil  \\
  \hline
  \rule[-5pt]{0pt}{15pt}
  \hfil $\bar{Q}_i$ \hfil 
& \hfil $\bar{\Box}$ \hfil 
  & \hfil $\bar{\Box} $ \hfil  \\
\hline
\end{tabular}\end{center}
\caption{Charge assignment for hidden matter fields in a model considered in Sec.~\ref{model}.
\label{table1}}
\end{table}

We assume 
that the SU($N$) gauge interaction becomes strong at low energy 
and the hidden-quarks are confined at the low-energy scale. 
This implies that 
the chiral symmetry is broken at the confinement scale, 
so that the low-energy effective theory can be described by 
pions $\pi_i$ ($i = 1, 2, 3, \dots , N_F^2 -1$) and baryons in the hidden sector. 
The mass of pions $m_\pi$ may be roughly given by 
\beq
 m_\pi \sim \sqrt{m_Q \Lambda}, 
 \label{m_pi}
\eeq
where $\Lambda$ is the dynamical scale of SU($N$). 
The hidden-pion decay constant $f_\pi$ is 
naively related to the dynamical scale of SU($N$) 
such as $4 \pi f_\pi / \sqrt{N} \approx \Lambda$. 
Note that the pions are stable due to the SU($N_F$)$_V$ flavour symmetry.

Once we omit the dynamics of the singlet field, 
our model is equivalent to the one discussed in Ref.~\cite{Hochberg:2014kqa}. 
They have found that 
the thermal relic density of pions can be consistent with the observed DM density 
and their self-interaction cross section can address the tension between 
astrophysical observations and $\Lambda$CDM model 
when the pion mass $m_\pi$ is about $100-500 \MeV$ 
and the pion decay constant $f_\pi$ is about $m_\pi /(5-10)$. 
Although their analysis is based on the chiral perturbation theory, 
the expansion parameter $m_\pi \sqrt{N} / (4 \pi f_\pi)$ is of order unity 
and the perturbation may break down. 
In fact, it has been discussed that the next-to-leading order effect becomes relevant 
in the interesting parameter region, 
so that there are $\order{1}$ uncertainties in their analysis~\cite{Hansen:2015yaa}. 
In this paper, 
we assume $m_\pi \sim 4 \pi f_\pi / \sqrt{N} \simeq \Lambda$ 
and use the naive dimensional analysis 
to estimate properties of pions, which we identify as self-interacting DM.

For the low energy effective theory, 
we write the effective lagrangian of pion fields by the naive dimensional analysis 
such as 
\beq 
 \mathcal{L}_\pi 
 &&= - \frac{1}{2} \Tr \lkk \del_\mu \pi \del^\mu \pi \rkk 
 - \frac{m_\pi^2}{2} \Tr \lkk \pi \pi \rkk 
 + \mathcal{L}_{2 \to 2} 
\nonumber
 \\
 && + c'_{\rm WZW} \frac{(4\pi)^3}{N^{3/2} \Lambda^5} 
 \epsilon^{\mu \nu \rho \sigma} \Tr \lkk \pi \del_\mu \pi \del_\nu \pi \del_\rho \pi \del_\sigma \pi \rkk 
 + \dots, 
\eeq
where $c'_{\rm WZW}$ is an $\order{1}$ parameter, 
$\mathcal{L}_{2 \to 2}$ represents terms that contribute to $\pi \pi \to \pi \pi$ 
scatterings, 
and the trace takes for the flavour indices. 
Terms with a odd number of pions are generically forbidden as long as CP-invarince is respected 
(we discuss the strong CP phase later). 
The forth term, however, respects the CP-invariance with a help of the anti-symmetric tensor 
$\epsilon^{\mu \nu \rho \sigma}$. 
When chiral symmetry is present, $c_{\rm WZW}$ has a quantization condition, 
but that is not essential here.
Note that it vanishes unless 
the number of pions $N_\pi$ is equal to or larger than five. 
Hereafter, we assume $N_F \ge 3$ ($N_\pi \ge 8$) 
and rewrite $c'_{\rm WZW}$ as $c_{\rm WZW} \equiv (m_\pi / \Lambda)^5 c'_{\rm WZW}$. 
We estimate the elastic scattering cross section for pions 
divided by $m_\pi$ 
such as 
\beq
 \frac{\sigma_{\rm ela}}{m_\pi} &=& \frac{(4 \pi)^4 c_1^2}{8 \pi N^2 m_\pi^3} 
 \nonumber \\
 &\simeq& 0.22 \cm^2 / g \ c_1^2 N^{-2} 
 \lmk \frac{m_\pi}{1 \GeV} \rmk^{-3}, 
 \label{sigma_ela}
\eeq
where we use the naive dimensional analysis 
and $c_1$ is an $\order{1}$ parameter. 
Observations of cluster collisions, including the bullet cluster, 
implies $\sigma_{\rm ela} / m_{\rm DM} \lesssim 0.47 \cm^2 / g$ 
and 
ellipticity on Milky way and cluster scales 
puts a constraint such as $\sigma_{\rm ela} / m_{\rm DM} \lesssim 0.5 \cm^2 / g$~\cite{Clowe:2003tk, Markevitch:2003at, Randall:2007ph, Rocha:2012jg, Peter:2012jh}. 
However, 
the cross section should be as large as this upper bound 
to address the "core-cusp" 
and "too-big-to-fail" problems~\cite{Spergel:1999mh, deBlok:2009sp, BoylanKolchin:2011de, Zavala:2012us}. 
Recently, it has been reported 
that there is an offset for the observation of Abell3827 cluster, 
which may be addressed by self-interacting DM 
with a cross section of $\sigma_{\rm ela} / m_{\rm DM} = 1.5 \cm^2 / g$~\cite{Kahlhoefer:2015vua}. 
These 
constraints and discussions 
have $\order{1}$ uncertainties 
due to, say, the difficulties of numerical simulations, 
so that 
in this paper we require that 
DM has self-interactions 
with the cross section of order $\sigma_{\rm ela} / m_{\rm DM} = 0.1-1 \cm^2 / g$ 
to address the above astrophysical problems. 
Equation~(\ref{sigma_ela}) 
then implies $m_\pi = \order{1} \GeV$, 
which is consistent with the original works within an $\order{1}$ uncertainty~\cite{Hochberg:2014kqa}.

Suppose that 
the hidden sector is in kinetic equilibrium 
with the SM sector, 
which is justified in the next subsection. 
When we can neglect $2 \to 2$ annihilation of pions, 
their Boltzmann equation can be written as~\cite{Hochberg:2014dra} 
\beq
 \dot{n}_\pi + 3H n_\pi = - \lmk n_\pi^3 - n_\pi^2 n_\pi^{\rm eq} \rmk 
 \la \sigma v^2 \ra_{3 \to 2}. 
 \label{3 to 2}
\eeq
Here, 
the thermal number density of $\pi$ is given by 
\beq
 n_\pi^{\rm eq} (T) &\simeq& N_\pi \lmk \frac{m_\pi T}{2 \pi} \rmk^{3/2} e^{- m_\pi / T}, 
\eeq
and the thermally-averaged 
cross section of $3 \to 2$ scattering process 
is calculated as~\cite{Hochberg:2014kqa}
\beq
 \la \sigma v^2 \ra_{3 \to 2} = 
 \frac{ (4 \pi)^6 c_{\rm WZW}^2 375 \sqrt{5}}{2 \pi N_F N^3 m_\pi^5} \frac{T^2}{m_\pi^2}, 
\eeq
where $T$ is a temperature of hidden sector 
and we assume $N_F \gg 1$ for simplicity. 
This implies that 
the pions freeze out at a temperature below its mass scale 
and the resulting relic density is given by 
\beq
 n_\pi^{\rm (FO)} 
 &\simeq& 
 \lmk \frac{3 H (T_F)}{\la \sigma v^2 \ra_{3 \to 2} } \rmk^{1/2}, 
\eeq
where $H(T_F)$ is the Hubble parameter at the time of pion freeze-out. 
As a result, we obtain 
\beq
 \frac{\rho_\pi}{s} 
 \simeq 
 0.2 \eV \ c_{\rm WZW}^{-1} N_F^{1/2} N^{3/2} 
 \lrfp{m_\pi}{1 \GeV}{3/2}, 
\eeq
where $T_F$ is the temperature at the pion freeze-out 
and we define $x_F \equiv m_\pi / T_F$ ($\simeq 24$). 
Thus, 
the pion abundance is consistent with the observed DM abundance of 
$\rho_{\rm DM}^{\rm (obs)} / s \simeq 0.4 \eV$ 
when $m_\pi = \order{1} \GeV$. 
In this parameter region, the self-interaction cross section of Eq.~(\ref{sigma_ela}) is consistent with 
the value indicated by astrophysical observations.

\subsection{Kinetic equilibrium between two sectors 
\label{equilibrium}}

Now let us check that 
the kinetic equilibrium between the hidden and SM sector 
is fulfilled until the pions freeze out. 
First, note that 
the singlet field $S$ interacts with the pions via a coupling of 
\beq
 \mathcal{L}_{\rm int} =  \frac{m_\pi^2}{2} \frac{S}{\la S \ra} \Tr \lkk \pi  \pi \rkk, 
 \label{s pi pi}
\eeq
where $\la S \ra$ is the VEV of $S$ at the vacuum. 
Note that the apparent singularity at $\la S \ra \to 0$ is cancelled by $m_\pi^2$ 
[see Eqs.~(\ref{singlet}) and (\ref{m_pi})]. 
Assuming a mixing between the singlet field $S$ and the SM Higgs field $h$, 
we can realize the kinetic equilibrium between the hidden and SM sectors. 
We denote the mixing parameter as $\theta$.

First, we consider the case of $r \equiv m_S / m_\pi \gg 1$, 
where $m_S$ is singlet mass. 
In this case, we can 
integrate out the singlet field to consider scatterings between pions and SM particles. 
Then we obtain the four-point interaction between the pions and electrons or muons 
by the mixing effect and the SM Yukawa interactions.%
\footnote{
The SM pions and photons can interact with the hidden pions 
and may contribute to their kinetic thermalization. 
We neglect it for simplicity because their contributions are the same order 
with that of muons. 
}
Here we explain the contribution of 
scatterings with muons 
because they dominate those with electrons for the case of $m_\pi \gtrsim 0.2 \GeV$. 
The cross section is roughly given by 
\beq
 \sigma_{\pi \mu \to \pi \mu} \approx 
 \frac{\theta^2 y_\mu^2}{8\pi} \frac{m_\pi^2 m_\mu^{2} }{\la S \ra^2 m_S^4}, 
\eeq
where $m_\mu$ ($\gtrsim T_F$) is the muon mass 
and 
$y_\mu$ is the muon Yukawa coupling. 
In order to be thermalized between successive $3 \to 2$ scatterings, 
pions need to lose kinetic energy about $m_\pi$ obtained from $3 \to 2$ scattering. 
Thus we estimate 
$\la \sigma v E_{\rm ex} / m_\pi \ra_{\pi \mu \to \pi \mu}  \approx \sigma_{\pi \mu \to \pi \mu} m_\mu / m_\pi$ 
where $E_{\rm ex}$ ($\approx m_\mu$) represents exchanged energy. 
In order to maintain the kinetic equilibrium between the two sectors at the pion freeze-out, 
we need to satisfy 
\beq
 \la \sigma v \frac{E_{\rm ex}}{m_\pi} \ra_{\pi \mu \to \pi \mu} 
 n_\mu^{\rm eq} (T_F) \gtrsim 
 \la \sigma v^2 \ra_{3 \to 2} \lmk n_\pi^{\rm eq} (T_F) \rmk^2 
 \simeq H(T_F). 
 \nonumber\\
\eeq
Here, the thermal number density of muons is given by 
\beq
 n_\mu^{\rm eq} (T) &\simeq& 4 \lmk \frac{m_\mu T}{2 \pi} \rmk^{3/2} e^{- m_\mu / T}. 
\eeq
Thus we have a constraint on $\theta$ depending on $m_\pi$ and $m_S$. 
We find that it could not be consistent with present constraints (discussed below) 
for the case of $r \gg 1$.

\begin{figure}[thhh]
\centering 
\includegraphics[width=.40\textwidth, bb=0 0 360 356]{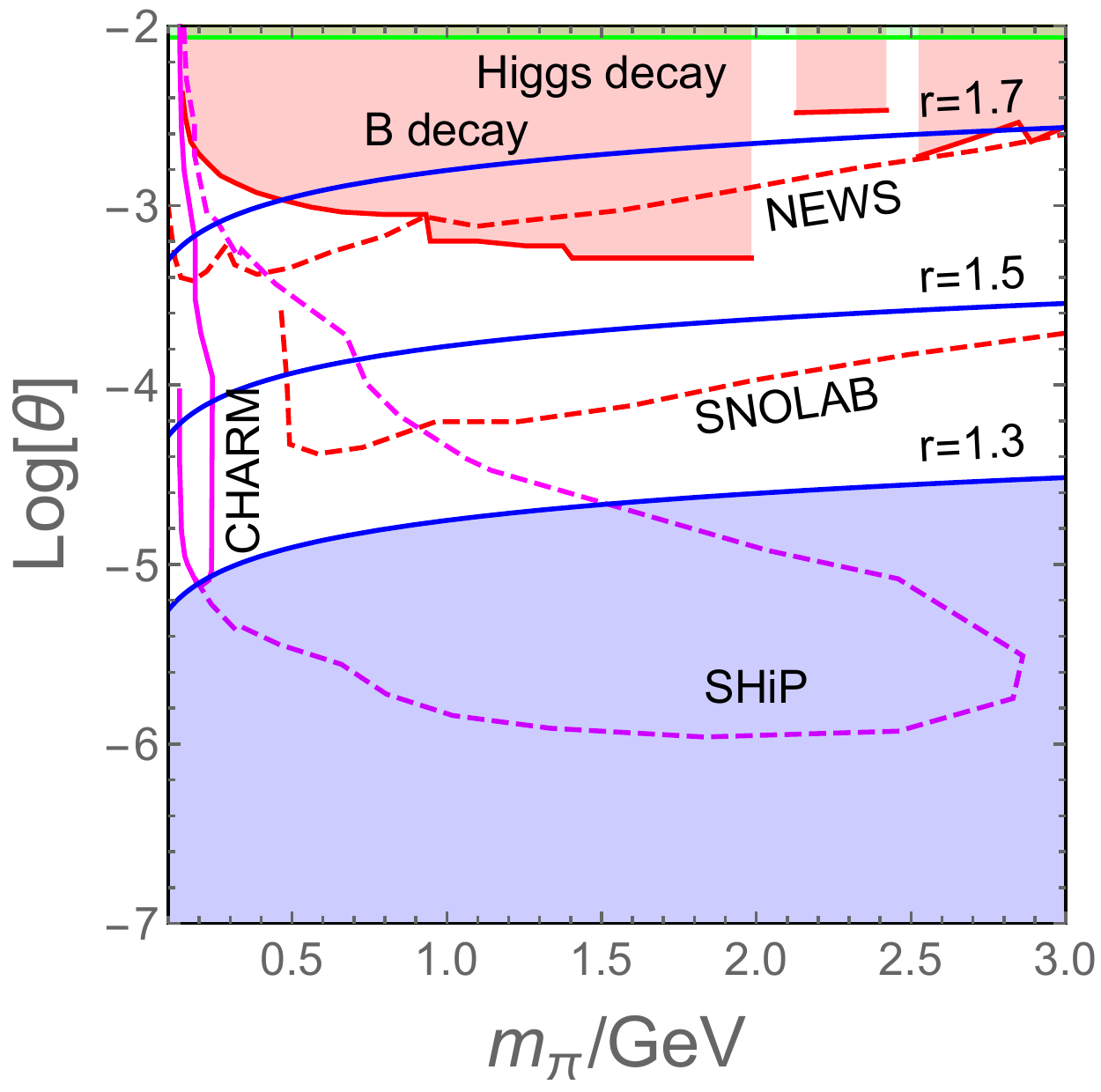} 
\\
\vspace{0.5cm}
\includegraphics[width=.40\textwidth, bb=0 0 360 356]{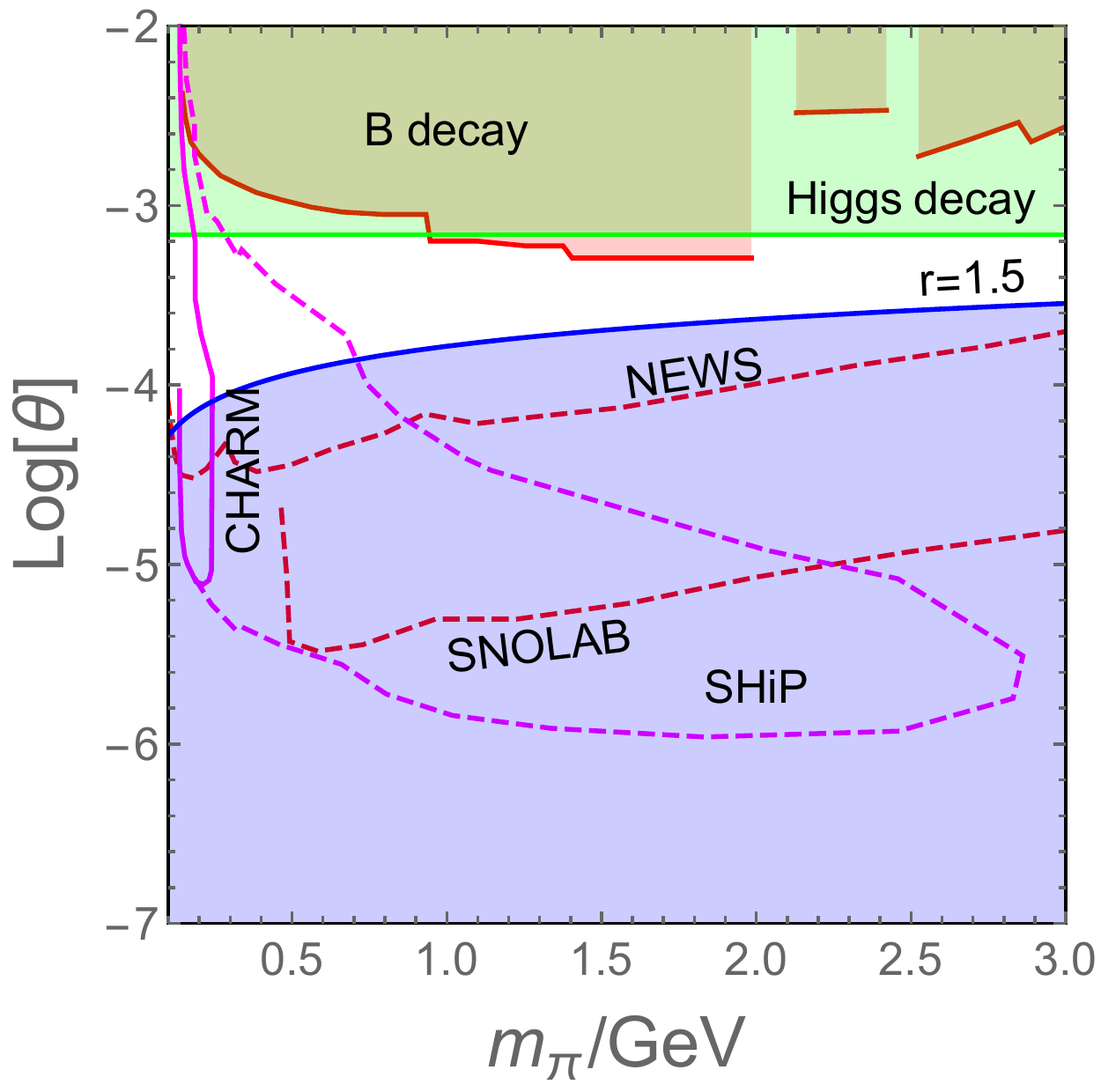} 
\caption{
Condition for kinetic equilibrium between the hidden and SM sectors for the case of $r \approx 1$. 
Above the blue curve for each value of $r$, 
we can maintain kinetic equilibrium. 
The green and red curves are the present upper bound on $\theta$ 
by the measurements of Higgs decay width~\cite{Khachatryan:2014iha, Aad:2015xua, Krnjaic:2015mbs} and B decay~\cite{Schmidt-Hoberg:2013hba}, respectively. 
The red-dashed curves 
are future sensitivities of DM direct detection experiments~\cite{Gerbier:2014jwa, Cushman:2013zza}. 
The region inside magenta curve 
is excluded by a beam-dump experiment~\cite{Bergsma:1985qz, Clarke:2013aya}, 
while that inside magenta-dashed curve 
is future sensitivity~\cite{Alekhin:2015byh}. 
To plot the constraints and sensitivities, 
we take $r = 1.5$ and $\lambda =1$ (upper panel) and $4 \pi$ (lower panel). 
}
  \label{fig1}
\end{figure}

Next, we consider 
the case that $m_S$ is larger than but close to $m_\pi$ 
($r \approx 1$). 
In this case, 
the number density of $S$ in the thermal plasma 
cannot be neglected at the time of pion freeze-out 
though it is suppressed by the Boltzmann factor. 
The singlet field $S$ elastically interacts with the pions 
with the cross section of order $\sigma_{S \pi \to S \pi} \approx m_\pi^2 / (8 \pi \la S \ra^4)$, 
where we use $m_S \simeq m_\pi$. 
The exchanged energy $E_{\rm ex}$ is now roughly given by $m_\pi$. 
Therefore, these particles are in kinetic equilibrium 
when 
\beq
 && \la \sigma v \frac{E_{\rm ex}}{m_\pi} \ra_{S \pi \to S \pi} n_S^{\rm eq} (T_F) \gtrsim 
 \la \sigma v^2 \ra_{3 \to 2} \lmk n_\pi^{\rm eq} (T_F) \rmk^2 
 \simeq H(T_F) 
 \nonumber \\
 \label{scattering rate1}
 \\
 &&\leftrightarrow 
 r \equiv \frac{m_S}{m_\pi} 
 \le r_{\rm max}, 
 \\
 &&r_{\rm max} \simeq 
 1+ \frac{1}{x_F} \ln \lkk \frac{\rho_\pi}{s} \frac{s(T_F)}{ m_\pi H(T_F)} \frac{\lambda^4}{8 \pi m_\pi^2}
\frac{1}{N_\pi} \lmk \frac{m_S}{m_\pi} \rmk^{3/2} \rkk 
 \nonumber\\
 &&\ \qquad \simeq  1+ \frac{1}{x_F} \ln \lkk 7.2 \times 10^6 N_\pi^{-1} r^{3/2}_{\rm max} \lambda^4 \lmk \frac{m_\pi}{1 \GeV} \rmk^{-2} \rkk, 
 \label{r max}
\eeq
where $n_S^{\rm eq} (T_F)$ is the thermal number density of $S$ at $T=T_F$: 
\beq
 n_S^{\rm eq} (T_F) &\simeq& \lmk \frac{m_S T_F}{2 \pi} \rmk^{3/2} e^{- m_S / T_F}. 
\eeq
Here we implicitly assume that $S$ is 
kinetically equilibrated with the SM sector via the decay and inverse-decay processes 
between successive $S \pi \to S \pi$ elastic scatterings. 
In fact, $S$ and $\pi$ are in kinetic equilibrium with the SM sector 
when \eq{scattering rate1} is satisfied 
and 
the decay and inverse decay rate of $S$ is larger than 
$H(T_F)  n_\pi^{\rm eq} / n_S^{\rm eq}$.%
\footnote{
This condition comes from the following discussion, for example. 
Suppose that the decay and inverse-decay of $S$ occur more rapidly than the $S \pi \to S\pi$ 
elastic scattering process (which rate is given by 
$\la \sigma v E_{\rm ex} / m_\pi \ra_{S \pi \to S \pi} n_\pi^{\rm eq}$). 
In this case, where the above condition follows from \eq{scattering rate1}, 
the kinetic equilibrium is reached between $\pi$ and $S$.
In the other case, i.e., 
if $S \pi \to S \pi$ elastic scattering process occurs more rapidly than the decay and inverse-decay of $S$, 
the energy of each $\pi$ is reduced by a factor of $n_\pi^{\rm eq} / (n_\pi^{\rm eq} + n_S^{\rm eq})$ 
and each $S$ obtains the energy of $E_\pi n_\pi^{\rm eq} / (n_\pi^{\rm eq} + n_S^{\rm eq})$. 
Here, $E_\pi$ is the initial energy of $\pi$ after every $3 \to 2$ process and is of order $m_\pi$, 
and $n_\pi^{\rm eq}$ is thermal number densities of $\pi$. 
Then, the energy density of $S$ is transferred to the SM plasma by its decay 
and inverse-decay processes. 
In every decay and inverse-decay process, 
the total energy density of $S$ and $\pi$ is reduced by a factor of $n_S E_\pi n_\pi^{\rm eq} / (n_\pi^{\rm eq} + n_S^{\rm eq})$. 
Therefore, in order to reduce the initial energy of $\pi$ by of order $E_\pi$ 
between successive $3 \to 2$ scatterings, 
the decay and inverse-decay rate should occur $(n_\pi^{\rm eq} + n_S^{\rm eq})/n_S^{\rm eq}$ ($\simeq n_\pi^{\rm eq} / n_S^{\rm eq}$) times. 
Finally we obtain the condition used in the main part of this paper. 
}
Since the decay rate of $S$ is proportional to $\theta^2$, 
we rewrite the latter condition as 
\beq
 \theta &\gtrsim& 
 \sqrt{\frac{8 \pi H (T_F)}{y_\mu^2 m_S} N_\pi  \lmk \frac{m_\pi}{m_S} \rmk^{3/2} e^{-m_\pi / T_F + m_S / T_F}} 
 \nonumber
 \\
 &\simeq& 
 2.3 \times 10^{-7} N_\pi^{1/2} r^{-5/4}  e^{x_F (r-1)/2} 
 \lmk \frac{m_\pi}{1 \GeV} \rmk^{1/2}. 
 \label{lower bound on theta}
\eeq
Thus, we have an upper bound on $r$ ($r_{\rm max}$) and a lower bound on $\theta$ 
to realize kinetic equilibrium between two sectors. 
The lower bounds of \eq{lower bound on theta} are plotted as blue curves in Fig.~\ref{fig1} 
for $r = 1.3, 1.5, 1.7$ ($r_{\rm max} \simeq 1.7$) and $\lambda = 1$ 
and 
for $r = 1.5$ ($r_{\rm max} \simeq 2.1$) and $\lambda = 4\pi$.

Here we should check that 
the annihilation of pions into singlet fields is inefficient 
at $T = T_F$ 
to justify the Boltzmann equation of Eq.~(\ref{3 to 2}). 
The annihilation rate is suppressed by a factor of $\exp \lkk - x_F (r-1) \rkk$ 
compared with the left-hand side of \eq{scattering rate1}. 
We find that 
the annihilation is inefficient when $r \gtrsim 1.1, 1.3, 1.5$ for $\lambda = 0.1, 1, 4 \pi$, 
respectively. 
Therefore, 
when 
$1.1 \lesssim r \lesssim 1.3$ for $\lambda = 0.1$, 
$1.3 \lesssim r \lesssim 1.7$ for $\lambda = 1$, 
and 
$1.5 \lesssim r \lesssim 2.1$ for $\lambda = 4 \pi$, 
the pions are in kinetic equilibrium with the SM sector 
and their annihilation can be neglected. 
For each $\lambda$, we have the lower bound on $r$, 
so that we cannot realize kinetic equilibrium in the lower-shaded region in Fig.~\ref{fig1}.

The nonzero mixing between the singlet field and the SM Higgs 
is constrained by the measurements of Higgs decay width at the LHC. 
The result is given by $\theta \lesssim 5.2 \times 10^{-2} /(N_F N \lambda)$~\cite{Khachatryan:2014iha, Aad:2015xua, Krnjaic:2015mbs}, 
which is shown as the horizontal green line in Fig.~\ref{fig1} 
for the case of $N_F =3$, 
$N=2$, $r=1.5$, 
and $\lambda = 1$ and $4 \pi$. 
Another constraint comes from $B$ decays at LHCb 
for $m_S \lesssim 5 \GeV$. 
The mixing parameter should be $\theta \lesssim (5-30) \times 10^{-4}$ for $m_S = \lkk 0.3, 5 \rkk \GeV$
though the regions $m_S = 2.95-3.18 \GeV$ and $m_S = 3.59-3.77 \GeV$ are vetoed 
in the experimental searches~\cite{Schmidt-Hoberg:2013hba}. 
It is plotted as the red curve in Fig.~\ref{fig1} for the case of $r=1.5$. 
Our predictions of mixing parameter 
for $r \lesssim 1.7$ 
are below the present upper bounds. 

International linear collider (ILC) experiment as well as LHC 
would precisely measure Higgs decay width, 
so that we would find a nonzero mixing between the hidden singlet field 
and the SM Higgs field. 
The mixing effect also leads to DM direct detection signals (see, e.g., Ref.~\cite{Krnjaic:2015mbs}). 
The present upper bound on the mixing parameter is larger than the above collider experiments~\cite{Angloher:2015ewa}, 
while 
some allowed region in Fig.~\ref{fig1} will be searched by future DM direct detection experiments, 
such as NEWS~\cite{Gerbier:2014jwa} and Super-CDMS SNOLAB~\cite{Cushman:2013zza}. 
Beam-dump experiments can search $S$ with a small mixing parameter. 
The CHARM collaboration puts a constraint for $m_\pi \lesssim 0.36  \GeV / r$~\cite{Bergsma:1985qz, Clarke:2013aya}, 
which is shown as magenta curves in the figures. 
We also plot future sensitivity of beam-dump experiment by the SHiP facility as magenta-dashed curve~\cite{Alekhin:2015byh}. 
The regions inside the magenta and magenta-dashed curves are excluded and will be proved 
by these beam-dump experiments, respectively. 
From Fig.~\ref{fig1}, 
we can see that 
a large parameter region will be searched in the near future.

Finally, let us comment on a strong CP phase in the hidden sector. 
The CP violating phase $\theta_{\rm CP}$ of SU($N$) interaction 
gives a term like $(4 \pi \theta_{\rm CP} m_\pi)/\sqrt{N} \Tr \lkk \pi \pi \pi \rkk$ in the low energy. 
It leads to unwanted $\pi + \pi \to \pi + S$ process, 
so that we have an upper bound on $\theta_{\rm CP}$ 
such as 
$\theta_{\rm CP} \lesssim 0.006 N^{1/2}/ \lambda$ for $r = 1.3$,
$\theta_{\rm CP} \lesssim 0.06 N^{1/2}/ \lambda$ for $r = 1.5$, 
and $\theta_{\rm CP} \lesssim 0.6 N^{1/2}/ \lambda$ for $r = 1.7$. 
We can explain such a small $\theta_{\rm CP}$ 
by forbidding the term by CP symmetry. 
Or, we can realize the Peccei-Quinn (PQ) mechanism in the hidden sector 
when we replace the singlet field $S$ with a complex scalar field 
with a global PQ symmetry~\cite{Peccei:1977hh, Peccei:1977ur}. 
After the PQ symmetry is spontaneously broken by the VEV of $S$, 
the CP violating phase is cancelled by the VEV of its phase component called axion. 
Note that when the axion mass is larger than the pion mass, 
we can neglect its effect on our analysis.

\subsection{Singlet potential 
\label{sec2-3}}

Here we explicitly write a model 
with a nonzero mixing between $S$ and $h$ as an example. 
We may write the potential of the singlet field 
and the SM Higgs field such as%
\footnote{
We define the origin of $S$ such that the mass of hidden quarks vanishes at $\la S \ra = 0$. 
}
\beq
 V (S, H) &&= 
 A_S S + \frac{1}{2} m_S^2 S^2 
 + \frac{1}{3} B_S S^3 
 + \frac{1}{4} \lambda_S S^4 
 \nonumber
 \\
 &&+ m' S \abs{H}^2 
 + \frac{g}{2} S^2 \abs{H}^2 
 + V (\abs{H}^2), 
\eeq
where 
$m_S$, $m'$, and $B_S$ are parameters with mass-dimension one, 
$g$ and $\lambda_S$ are dimention less parameters, 
$A_S$ is a parameter with mass-dimension three, 
and $V( \abs{H}^2)$ is the Higgs potential. 
The singlet field $S$ acquires 
a nonzero VEV due to the first and the third terms. 
Note that 
we need $\la S \ra = m_Q / \lambda$. 
After $H$ obtains a VEV at the electro-weak phase transition, 
the fifth and sixth terms lead to a mixing between $S$ and the SM Higgs field 
such as $\theta \simeq (m' v_h + g \la S \ra v_h) / (2 m_h^2)$, 
where $v_h$ ($\simeq 246 \GeV$) and $m_h$ ($\simeq 125 \GeV$) are the Higgs VEV and mass, respectively. 
We can obtain a small but nonzero mixing 
that is consistent with 
\eq{lower bound on theta}.

\section{Model with strong U(1) 
\label{model2}}

In this section, we consider another SIMP model 
in a strongly-interacting Abelian gauge theory. 
We consider a hidden Abelian gauge theory with a scalar monopole $\phi$ 
and $N_F$ pairs of hidden electrons and positrons $\psi_i$ and $\bar{\psi}_i$. 
The charge assignment for the hidden electrons and positrons are shown in Table~\ref{table2}. 
We denote the electric coupling 
as $g_e$ 
and the magnetic coupling as $g_m$, 
which satisfy Dirac quantization condition: $g_e g_m = 2 \pi n$ ($n = 1,2,3, \dots$). 

As we see below, the monopole develops condensation at low-energy scale 
to confine the U(1)$_H$ gauge interaction. 
Its radial component has a mass of order the confinement scale 
and can mix with the Higgs field. 
In this model, therefore, 
we do not need to introduce the singlet field to mediate two sectors.

\begin{table}\begin{center}
\begin{tabular}{|p{1.0cm}|p{1.5cm}|p{1.5cm}|p{1.5cm}|p{1.5cm}|}
  \hline
  \rule[-5pt]{0pt}{15pt}
& \hfil SU($N_F$)$_V$ \hfil 
    & \hfil U(1)$_H$ \hfil & \hfil U(1)$_Y$ \hfil  \\
  \hline
  \rule[-5pt]{0pt}{15pt}
  \hfil $\psi_i$ \hfil 
& \hfil $\Box$ \hfil 
  & \hfil $1$ \hfil & \hfil 0 \hfil  \\
  \hline
  \rule[-5pt]{0pt}{15pt}
  \hfil $\bar{\psi}_i$ \hfil 
& \hfil $\bar{\Box}$ \hfil 
  & \hfil $-1$ \hfil & \hfil 0 \hfil  \\
\hline
\end{tabular}\end{center}
\caption{Charge assignment for hidden matter fields in a model considered in Sec.~\ref{model2}.
\label{table2}}
\end{table}

\subsection{Monopole condensation and mixing with Higgs
\label{model2}}

We write the potential of scalar monopole such as 
\beq
  V(\phi) = - \mu^2 \abs{\phi}^2 + \lambda_\cphi \abs{\phi}^4, 
  \label{potential}
\eeq
where $\lambda_\cphi$ is expected to be of order $(4\pi)^2$ by the naive dimensional analysis.%
\footnote{
Strictly speaking, 
it is not possible to write a Lagrangian for the system of both electrons and monopoles, 
and our equations like \eq{potential} should be regarded 
as a schematic picture of what is going on rather than a precise equation. 
Naive dimensional analysis should be applied to physical quantities like masses of particles 
and scattering amplitudes rather than ill-defined ``parameters in the Lagrangian''.
}
At the minimum of the potential, 
the monopole develops a condensation such as $\sqrt{2} \la \abs{\phi} \ra = \mu / \sqrt{\lambda_\cphi}$ 
($\equiv v$). 
Then, the mass of radial component of monopole, which we denote as $\cphi$ ($\equiv \sqrt{2} \abs{\phi}$), 
is given by $m_\cphi = \sqrt{2} \mu$. 
We expect that the mass of hidden \UH gauge boson 
$m_v$ is of order $m_\cphi$. 
Hereafter, we assume $m_v$ to be larger than $m_\pi$ and neglect its effect except for in Sec.~\ref{sec:kinetic mixing}. 
After the monopole acquires the VEV, 
hidden electrons are attached by strings via the Meisner effect 
and are 
confined by the tension of the string~\cite{Nambu:1974zg}. 
Its tension $\mu_s$ determines the dynamical scale and is given as 
\beq
 \mu_s = 
\frac{g_e^2 g_m^2}{8 \pi} 
 v^2 \log \lmk \frac{m_\cphi^2}{m_v^2} + 1 \rmk, 
\eeq
which is almost independent of $g_e$ and $g_m$ due to the Dirac quantization condition. 

We have composite particles $\pi_i$ below the confinement scale. 
There is no baryon state at low energy 
because 
baryons cannot be neutral under U(1)$_H$. 
Although in the previous section we add a singlet field $S$ to maintain 
the kinetic equilibrium between the hidden and SM sectors, 
we can realize it via the mixing between the monopole and the SM Higgs field 
without adding the singlet field. 
We do not assume chiral symmetry in the hidden sector, 
which implies that pions have a mass of order the dynamical scale $m_\pi \approx \Lambda$.%
\footnote{
Or we can just write an electron mass term to make pions massive. 
If the electron mass $m_\psi$ is smaller than $\Lambda$, 
we might have an approximate chiral symmetry that is expected to be 
dynamically broken by the electron confinement. 
In this case, we have $m_\pi \simeq \sqrt{m_\psi \Lambda}$. 
}
Note that we assume SU($N_F$)$_V$ flavour symmetry to make pions stable (see Table.~\ref{table2}).

\begin{figure}[t]
\centering 
\includegraphics[width=.40\textwidth, bb=0 0 360 356]{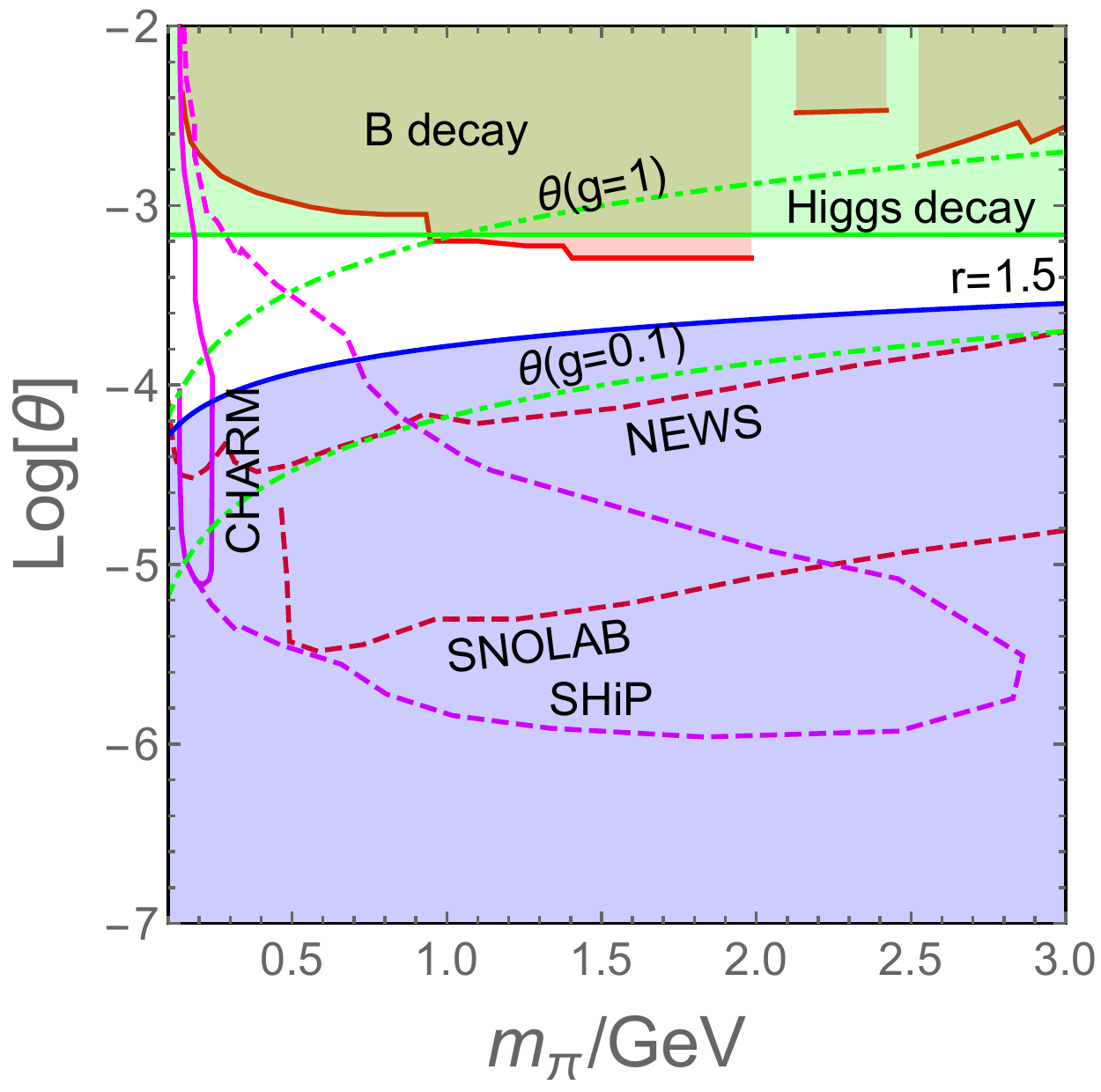} 
\caption{
Same as Fig.~\ref{fig1} but for the model considered in Sec.~\ref{model2}. 
The green dot-dashed curves represent the values of $\theta$ 
for $g=0.1$ and $1$. 
}
  \label{fig2}
\end{figure}

Below the confinement scale, 
we have an interaction between the radial component of monopole $\cphi$ and pions 
such as 
\beq
 \mathcal{L} = c_\cphi \frac{m_\pi^2}{2} \frac{\cphi}{v} \Tr \lkk \pi \pi \rkk, 
 \label{cphi pi pi}
\eeq
where $c_\cphi$ is an $\order{1}$ constant. 
Hereafter we take $c_\cphi = 1$. 
We introduce the following term to obtain the mixing between 
$\cphi$ and $h$:%
\footnote{
The coupling constant $g$ may be smaller than $\order{1}$ 
due to an anomalous dimension of monopole 
because our model may be conformal above the monopole and electron mass scale. 
Unfortunately, we cannot determine the anomalous dimension of the monopole. 
}
\beq
 V_{\rm int} = g \abs{\phi}^2 \abs{H}^2, 
\eeq
This term gives a mixing between $\cphi$ and $h$ 
with $\theta \simeq g v_h v / 2 m_h^2$. 
Noting that $\lambda_\cphi \approx (4 \pi)^2$, 
we find 
\beq
 \theta \simeq 4.4 \times 10^{-4} g \lmk \frac{m_\cphi}{1 \GeV} \rmk. 
\eeq
Replacing $m_S$ with $m_\cphi$ and $\la S \ra$ with $v$, 
we can quote the calculations in Sec.~\ref{equilibrium}. 
The result is shown in Fig.~\ref{fig2}, 
where the hidden sector can be in kinetic equilibrium with the SM sector above the blue curve. 
The parameter $\theta$ is plotted as green dot-dashed curves, 
where we take $g=0.1$ and $1$. 
Testabilities and constraints of this model are the same as the ones considered in the previous section 
(see the last two paragraphs in Sec.~\ref{equilibrium}) 
except for an additional signal explained in the next subsection.

\subsection{Kinetic mixing\label{sec:kinetic mixing}}

Here we comment on a kinetic mixing between 
the U(1)$_H$ and U(1)$_Y$ gauge bosons~\cite{Yamada:2016jgg}. 
First, 
note that 
the Abelian gauge theory 
may be conformal in the presence of monopole as well as electrons. 
In this case, its gauge field strength $F_{\mu \nu}$ has an scaling dimension larger than $2$, 
which is guaranteed by the unitarity bound~\cite{Mack:1975je}. 
This implies that 
the kinetic mixing term $\chi B_{\mu \nu} F^{\mu \nu}$ is an irrelevant operator 
and is suppressed at low energy. 
The resulting kinetic mixing parameter $\chi$ 
at low energy 
depends on the value of anomalous dimension of the U(1)$_H$ gauge boson $F^{\mu \nu}$, 
though we do not have information about it. 
Thus our model predicts 
a very small kinetic mixing between U(1)$_H$ and U(1)$_Y$ gauge bosons.

Once there is a mixing between U(1)$_H$ and U(1)$_Y$ gauge bosons, 
we may expect a term like%
\footnote{
In the presence of monopole, 
the gauge field strength does not satisfy the Bianchi identity. 
As a result, we have no reason 
that we can omit the term of 
\eq{kinetic mixing}. 
If $F_{\mu \nu}$ satisfies Bianchi identity and thus the gauge field is well-defined, 
in the mass eigenstate of gauge fields, only the Z-boson couples to 
$\epsilon^{\mu \nu \rho \sigma}  \Tr \lkk \pi \del_\rho \pi \del_\sigma \pi \rkk$ at the tree level. 
Then, \eq{kinetic mixing} is expected to have a small coefficient (if exist) in Refs.~\cite{Lee:2015gsa, Hochberg:2015vrg}. 
If Bianchi identity is not satisfied, 
operator mixing by the strong dynamics between 
$F^{\mu \nu}$ and $\epsilon^{\mu \nu \rho \sigma} \Tr \lkk \pi \del_\rho \pi \del_\sigma \pi \rkk$ 
is not forbidden 
(or in other words, $F^{\mu \nu}$ can create/annihilate three pion states as well as one vector boson states) 
and 
\eq{kinetic mixing} is generated by replacing $F^{\mu \nu}$ 
by $\epsilon^{\mu \nu \rho \sigma} \Tr \lkk \pi \del_\rho \pi \del_\sigma \pi \rkk$ in $\chi B_{\mu \nu} F^{\mu \nu}$. 
}
\beq
 \mathcal{L} \supset \frac{(4\pi)^2}{\Lambda^3} \chi \epsilon^{\mu \nu \rho \sigma} 
B_{\mu \nu} \Tr \lkk \pi \del_\rho \pi \del_\sigma \pi \rkk, 
\label{kinetic mixing}
\eeq
where we omit an $\mathcal{O}(1)$ uncertainty factor. 
This term leads to unwanted $\pi + \pi \to \pi + \gamma$ annihilation process, 
so that its cross section should be suppressed such as 
\beq
 \la \sigma v \ra_{\pi \pi \to \pi \gamma} n_\pi^{\rm eq} (T_F) &\lesssim& 
 \la \sigma v^2 \ra_{3 \to 2} \lmk n_\pi^{\rm eq} (T_F) \rmk^2 
 \simeq H(T_F), 
 \nonumber
 \\
\eeq
where a rough estimation gives 
\beq
 \la \sigma v \ra_{\pi \pi \to \pi \gamma}  &\sim& 
 \chi^2 \frac{(4\pi)^4}{8\pi m_\pi^2} \lmk \frac{T}{m_\pi} \rmk. 
\eeq
This can be rewritten as 
\beq
 \chi \lesssim 6 \times 10^{-5} 
 \lmk \frac{m_\pi}{1 \GeV} \rmk, 
 \label{lower bound on chi} 
\eeq
within an $\mathcal{O}(1)$ uncertainty. 
Such a small kinetic mixing parameter is consistent with our model 
because $\chi B_{\mu \nu} F^{\mu \nu}$ is an irrelevant operator 
and is suppressed at low energy as explained above. 

The kinetic mixing is constrained by many experiments (see Refs.~\cite{Lee:2015gsa, Hochberg:2015vrg}). 
A model-independent bound comes from 
electroweak precision tests 
because nonzero kinetic mixing modifies 
parameters in the EW sector. 
They put an upper bound such as $\chi \lesssim 2 \times 10^{-2}$ 
for $m_v \lesssim \order{1} \GeV$~\cite{Essig:2013vha}. 
In our model, 
the massive hidden photon dominantly decays into hidden pions for $m_v \gtrsim 2 m_\pi$, 
in which case 
BaBar experiment puts a constraint 
as $\chi \lesssim 10^{-3}$~\cite{Lees:2014xha}. 
Even for $m_v \lesssim 2 m_\pi$, 
it puts a similar constraint~\cite{Aubert:2008as}. 
In the near future, 
Bell-II experiment can measure the mixing parameter of order $10^{-4}$~\cite{Essig:2013vha, Soffer:2014ona}. 
We may not expect 
that we can observe kinetic mixing effect suppressed by \eq{lower bound on chi}.

\section{Discussion and conclusions
\label{conclusions}}

We have provided self-interacting DM models 
that explain the discrepancy between astrophysical observations and $\Lambda$CDM model. 
The models are based on low-energy effective theories of hidden QCD, 
where hidden pions are identified as SIMP DM. 
The thermal relic abundance of DM is determined by $3 \to 2$ scattering process 
and is consistent with the observed DM abundance as discussed in Ref.~\cite{Hochberg:2014kqa}. 
We have first investigated a non-Abelian gauge theory with a singlet field. 
The condition for kinetic equilibrium between the hidden and SM sectors 
can be realized by a mixing between the singlet field and the SM Higgs field. 
The nonzero mixing parameter will be measured by future experiments, 
such as LHC, ILC, NEWS~\cite{Gerbier:2014jwa}, Super-CDMS SNOLAB~\cite{Cushman:2013zza}, and SHiP~\cite{Alekhin:2015byh}. 

Then we have provided a composite SIMP model originating from 
U(1)$_H$ confinement due to monopole condensation. 
In this case, 
the radial component of monopole can mix with the SM Higgs field, 
so that it mediate the hidden and SM sectors without introducing the additional singlet field. 
It is outstanding that 
the monopole plays the roles of U(1)$_H$ confinement and mediator between two sectors. 
In addition, 
there is no unwanted baryon in this theory because baryons are charged under U(1)$_H$.

\vspace{1cm}

%
\section*{Acknowledgments}
T.~T.~Y thanks H. Murayama for useful discussion. 
A. K. would like to acknowledge the Mainz institute for Theoretical Physics (MITP) 
for its hospitality and its partial support during the completion of this work.
This work is supported by Grant-in-Aid for Scientific Research 
from the Ministry of Education, Science, Sports, and Culture
(MEXT), Japan, 
No. 26104009 and No. 26287039 (T.T.Y), 
World Premier International Research Center Initiative
(WPI Initiative), MEXT, Japan, 
and the JSPS Research Fellowships for Young Scientists (M.Y.). 
%

\vspace{1cm}



\end{document}